# Benchmark Problems for Constraint Solving

**Alin Suciu, Rodica Potolea, Tudor Mureşan**

Department of Computer Science

Technical University of Cluj-Napoca

26-28, George Bariţiu St., RO-3400, Cluj-Napoca, Romania

Tel/Fax: +40-64-194491

{Alin.Suciu, Rodica.Potolea, Tudor.Muresan}@cs.utcluj.ro

**Abstract:** Constraint Programming is roughly a new software technology introduced by Jaffar and Lassez in 1987 for description and effective solving of large, particularly combinatorial, problems especially in areas of planning and scheduling. In the following we define three problems for constraint solving from the domain of electrical networks; based on them we define 43 related problems. For the defined set of problems we benchmarked five systems: ILOG OPL, AMPL, GAMS, Mathematica and UniCalc. As expected some of the systems performed very well for some problems while others performed very well on others.

## 1 Introduction

Constraint Programming is roughly a new software technology introduced by Jaffar and Lassez in 1987 for description and effective solving of large, particularly combinatorial, problems especially in areas of planning and scheduling. Constraint Programming aims at developing techniques for efficiently search for solutions in the possible solution space, usually a very large one. In constraint satisfaction problems the programming process consists of the generation of requirements (constraints) and solution of these requirements, by specialized constraint solvers. It is basically intended to complete Mathematical Programming rather than to replace it. Constraints can deal with partial information, and specialized techniques take advantage of this property. Currently the domain may be split into Constraint Satisfaction and Constraint Solving. It is a very dynamic field of research today and there is a lot of ongoing research, therefore we may expect some important results in the near future.

In the following we define two benchmark problems for constraint solving from the domain of electrical networks. Other related problems can easily be generated by some simple modifications.

One of the problems uses linear constraints only, the other one uses diodes as example of non-linear behavior – however, the diode model used here is a piece-wise linear model, i.e. a composition of two linear operational modes. Both problems are scalable with respect to the number of electrical components and hence with respect to the number of variables and constraints. Exact solutions can easily be obtained by, more or less, basic computation (we are going to give the exact expected solutions in order to ease the verification of the accuracy with which the constraint solvers actually solve the problems).

A further dimension on which different test instances can be generated refers to the precision with which the resistor values are defined (i.e. real numbers or intervals). This checks the ability of the constraint solvers to deal with uncertain knowledge in the form of intervals.

## 2   The Benchmark Examples

We can now, by depicting electrical networks using the above icons, specify sets of and-or-connected constraints which must simultaneously hold in order to describe the behavior of the whole electrical network.

We will present below the following benchmarks:
- the baby example (BE)
- the first benchmark example (FB)
- the second benchmark example (SE)

The baby example aims at showing a very simple example of electric circuit modeling with constraints while the first and the second examples are quite close to real problems and are more difficult to solve.

### *2.1   The Baby Example*

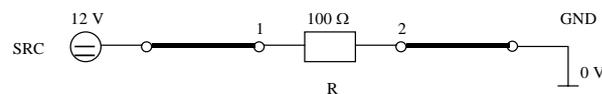

**Figure 1: A „Baby Example" with just one resistor**

The baby example in Figure 1 with the components „SRC", „R", and „GND", and two wires defines the constraint set:

```
u_SRC = 12;                                    (* source *)
i_SRC + i1_R = 0; u_SRC = u1_R;                (* wire *)
i1_R + i2_R = 0; u1_R - u2_R = 100 * i1_R;     (* resistor R=100 *)
i2_R + i_GND = 0; u2_R = u_GND;                (* wire *)
u_GND = 0.                                     (* ground *)
```

### *2.2   The First Example*

The problem (i.e. problem family) is given in Figure 2.

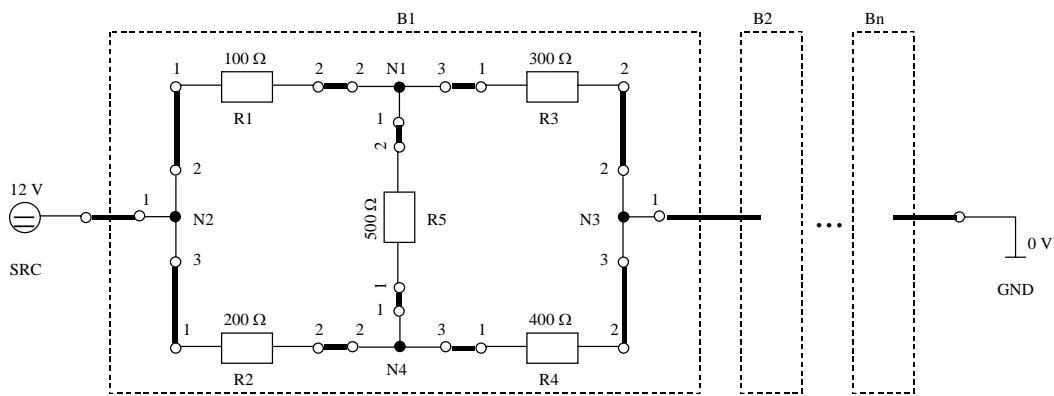

**Figure 2: The First Benchmark Example**

Note that all the dotted boxed shall contain the same sub-structure. So there are basically n boxes in the network, all of them having the same structure and resistance values (i.e. as the left-most box). This gives us a constraint problem with 2*totalNumberOfPorts = 2*(22n+2) = 44n+4 variables, and 42n+2+2+(n-1)*2+2 = 44n+4 constraints.

## *2.3 The Second Example*

The second family of problems is similar to the first one, but replaces in all boxes the resistors R1 and R4 with diodes D1 and D4, as in Figure 3. The remaining resistors are assumed to be 100 Ω (or some interval around that value).

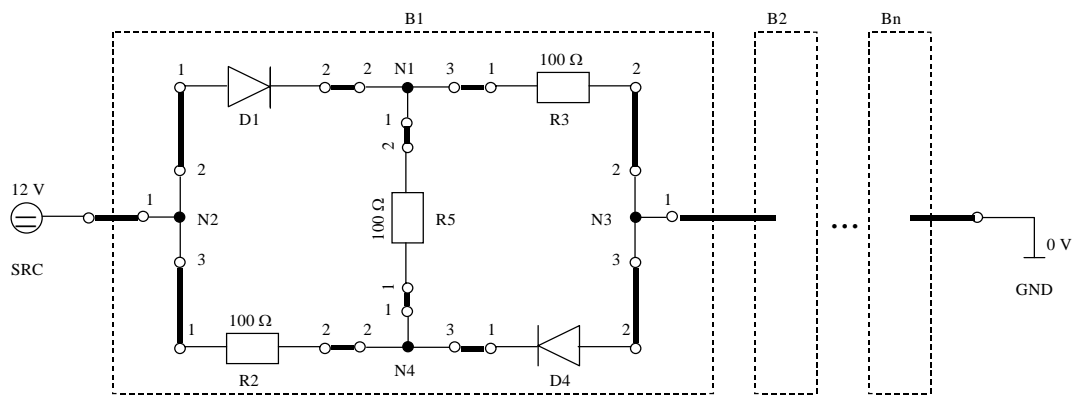

**Figure 3: The Second Benchmark Example**

After computing the circuit there is only one non-zero current in the network that takes: the upward direction in the left Kirchhoff node, the straight direction in the top node, and the box-leaving direction in the right node of each box.

Note that, because each diode introduces a disjunction, there are in the worst case (for n boxes) $4^n$ alternative linear constraint sets that have to be considered by the solver. There are 44n+4 variables in the problem with n boxes and 44n+4 constraints in each alternative branch containing only linear constraints.

However, the problem is not as hard as the above number of branches might suggest (the number of branches that have to be investigated seems to be limited, so an intelligent solver should get linear solving times in n).

# 3 Solving the Problems

There are several things we might want from a solver in the context of the above examples:

1) solving for the ground current
    a) fixed values - we want the value of the ground current i_GND, as a fixed value, given the fixed values of the resistors and source voltage
    b) interval values - we want to be able to specify the value of a resistor with some tolerance, i.e. as an interval and obtain the ground current as an interval
2) numerical constraint solving
    a) fixed values - we want the fixed values of all the unknowns in the constraint set given fixed input
    b) interval values - we want the interval values of all the unknowns in the constraint set given interval input
3) symbolic constraint solving - we want the symbolic expression of all the unknown in terms of the known input
4) diagnosis as optimization - we want to be able to maximize (minimize) an objective function expressing the probability of the correct (faulty) behavior of some components of the circuit

## 3.1 Solving the Baby Example (BE)

The following constraint system (system of equations) describes the Baby Example; the values of the source voltage and the resistance are substituted in the equations and the system is linear.

```
u_SRC = 12;
i_SRC + i1_R = 0;
u_SRC = u1_R;
i1_R + i2_R = 0;
u1_R - u2_R = 100 * i1_R;
i2_R + i_GND = 0;
u2_R = u_GND;
u_GND = 0;
```

The numerical solution using fixed values is:

```
u_SRC = 12.00000000                    u1_R = 12.00000000
i_SRC = -0.12000000                    u2_R = 0.00000000
i1_R = 0.12000000                      i_GND = 0.12000000
i2_R = -0.12000000                     u_GND = 0.00000000
```

The symbolic solution using fixed values is: `i_GND = u_SRC / R`.

Various features of the benchmarked systems can be evaluated using only this small example, like their ability to deal with interval computations or symbolic computations. The following problems can be generated based on this example.

### 3.1.1 BLNS. Linear Numerical Solving

**BLNS1**. Solve the system for fixed values of the voltage and the resistance. The resistance is given as a fixed value, and the values of the currents and voltages should be obtained as fixed values.
**BLNS2**. Solve the system for interval value of the resistance (tolerance). The resistance is given as an interval, and the values of the currents and voltages should be obtained as intervals.
**BLNS3**. Solve the system for alternate fixed values of the resistance. The resistance is given as either fixed value R1 OR fixed value R2, and the fixed values of the currents and voltages should be obtained as alternate fixed values corresponding to the two values of the resistance.
**BLNS4**. Solve the system for alternate interval values of the resistance (tolerance). The resistance is given as either interval value R1 OR interval value R2, and the interval values of the currents and voltages should be obtained as alternate interval values corresponding to the two values of the resistance.

### 3.1.2 BNNS. Nonlinear Numerical Solving

We modify the initial system replacing the equation: `u1_R - u2_R = 100 * i1_R;` with two equations, as follows:
```
u1_R - u2_R = R * i1_R; R = 100;
```
The idea here is to see if a system is capable of reducing the resulting nonlinear system to a linear one and solve it.

**BNNS1**. Solve the system for fixed values of the voltage and the resistance. The resistance is given as a fixed value, and the values of the currents and voltages should be obtained as fixed values.
**BNNS2**. Solve the system for interval value of the resistance (tolerance). The resistance is given as an interval, and the values of the currents and voltages should be obtained as intervals.

**BNNS3**. Solve the system for alternate fixed values of the resistance. The resistance is given as either fixed value R1 OR fixed value R2, and the fixed values of the currents and voltages should be obtained as alternate fixed values corresponding to the two values of the resistance.

**BNNS4**. Solve the system for alternate interval values of the resistance (tolerance). The resistance is given as either interval value R1 OR interval value R2, and the interval values of the currents and voltages should be obtained as alternate interval values corresponding to the two values of the resistance.

### 3.1.3 BSS. Symbolic Solving

We modify the initial system by removing the equation: `u_SRC = 12;` and replacing the equation: `u1_R - u2_R = 100 * i1_R;` with `u1_R - u2_R = R * i1_R;`

**BSS1**. Solve the system for fixed values of the voltage and the resistance. The resistance is given as a fixed value, and the values of the currents and voltages should be obtained as fixed values.

**BSS2**. Solve the system for interval value of the resistance (tolerance). The resistance is given as an interval, and the values of the currents and voltages should be obtained as intervals.

**BSS3**. Solve the system for alternate fixed values of the resistance. The resistance is given as either fixed value R1 OR fixed value R2, and the fixed values of the currents and voltages should be obtained as alternate fixed values corresponding to the two values of the resistance.

**BSS4**. Solve the system for alternate interval values of the resistance (tolerance). The resistance is given as either interval value R1 OR interval value R2, and the interval values of the currents and voltages should be obtained as alternate interval values corresponding to the two values of the resistance.

### 3.1.4 BLO. Linear Optimization

**BLO1**. Simple linear optimization. With the initial system of equations, minimizing and maximizing some variable should give an identical result, the same result as BLNS1.

**BLO2**. Support for interval computations and linear optimization. We want to be able to directly specify the resistance as an interval and to minimize/maximize some variable (e.g. i_GND).

**BLO3**. Interval computation simulation. If a system does not allow interval computation, we can simulate this by means of optimization replacing the equation `u1_R - u2_R = 100 * i1_R;` with two inequalities: `u1_R - u2_R >= 90 * i1_R; u1_R - u2_R <= 110 * i1_R;` and then minimizing / maximizing the desired variable (e.g. i_GND).

**BLO4**. Strict inequalities. Replacing the nonstrict inequalities from BLO3 with strict ones, we can check the behavior of a system with respect to strict inequalities.

**BLO5**. Direct support for OR-ed constraints in linear optimization. By replacing the equation `u1_R - u2_R = 100 * i1_R;` with the equation: `u1_R - u2_R = 90 * i1_R or u1_R - u2_R = 110 * i1_R;` and minimizing/maximizing some variable we can check if a system supports OR-ed constraints. We should obtain results similar to BLNS3.

**BLO6**. Simulation of OR-ed constraints. If a system does not support OR-ed constraints directly, we can try to simulate them by means of binary variables as follows:
```
binary variable a,b;
a = (u1_R - u2_R = 90 * i1_R);
b = (u1_R - u2_R = 110 * i1_R);
a + b >= 1;
```

**BLO7**. Diagnosis as optimization. This requires either direct or simulated OR-ed constraints. We replace the equation describing the behavior of the resistor (`u1_R - u2_R = 100 * i1_R;`) with a disjunction between the correct and the faulty behavior:`(p=0.9 and u1_R - u2_R = 100 * i1_R) or (p=0.1 and i1_R=0);` where p is the probability of the correct / faulty behavior.

Maximizing p in the resulting system must give the same results as BL1 and the value of p must be 0.9. Adding the measured constraint i2_R = 0 must give the value of p = 0.1.

### 3.1.5 BNO. Nonlinear Optimization

We modify the initial system replacing the equation: `u1_R - u2_R = 100 * i1_R;` with two equations, as follows:

```
u1_R - u2_R = R * i1_R; R = 100;
```

The idea here is to see if a system is capable of dealing with the resulting nonlinear system and perform optimization on some variable (e.g. i_GND).

**BNO1**. Simple nonlinear optimization. With the resulting system of equations, minimizing and maximizing some variable should give an identical result, the same result as BNNS1.
**BNO2**. Support for interval computations and nonlinear optimization. We want to be able to directly specify the resistance as an interval and to minimize/maximize some variable (e.g. i_GND).
**BNO3**. Interval computation simulation. If a system does not allow interval computation, we can simulate this by means of optimization replacing the equation `u1_R - u2_R = R * i1_R;` with two inequations:
```
u1_R - u2_R >= 0.90 * R * i1_R;
u1_R - u2_R <= 1.10 * R * i1_R;
```
and then minimizing / maximizing the desired variable (e.g. i_GND).
**BNO4**. Direct support for OR-ed constraints in nonlinear optimization. By replacing the equation `u1_R - u2_R = 100 * i1_R;` with the equation: `u1_R - u2_R = 0.90 * R * i1_R or u1_R - u2_R = 1.10 * R * i1_R;` and minimizing/maximizing some variable we can check if a system supports OR-ed constraints. We should obtain results similar to BLO5.

## 3.2 Solving the First Example (FE)

### 3.2.1 FEOB. One Box

The constraint system (system of equations) describing the first example with just one box is:

```
resistors:
   for j=1 to 5
      i1_Rj + i2_Rj = 0; u1_Rj - u2_Rj = Rj * i1_Rj;
nodes:
   for j=1 to 4
      i1_Nj + i2_Nj + i3_Nj = 0; u1_Nj = u2_Nj; u1_Nj = u3_Nj;
wires:
      i1_N1 + i2_R5 = 0; u1_N1 = u2_R5;
      i2_N1 + i2_R1 = 0; u2_N1 = u2_R1;
      i3_N1 + i1_R3 = 0; u3_N1 = u1_R3;
      i2_N2 + i1_R1 = 0; u2_N2 = u1_R1;
      i3_N2 + i1_R2 = 0; u3_N2 = u1_R2;
      i2_N3 + i2_R3 = 0; u2_N3 = u2_R3;
      i3_N3 + i2_R4 = 0; u3_N3 = u2_R4;
      i1_N4 + i1_R5 = 0; u1_N4 = u1_R5;
      i2_N4 + i2_R2 = 0; u2_N4 = u2_R2;
      i3_N4 + i1_R4 = 0; u3_N4 = u1_R4;
source:
      u_SRC = 12;
ground:
      u_GND = 0;
SRC & GND wires:
      i_SRC + i1_N2 = 0; u_SRC = u1_N2;
      i1_N3 + i_GND = 0; u1_N3 = u_GND;
```

The numerical solution using fixed values is:

| | | | |
|---|---|---|---|
| iR[1,1] = 0.03105882 | uR[1,3] = 8.89411765 | iN[2,1] = 0.03105882 | uN[2,1] = 8.89411765 |
| iR[1,2] = 0.01905882 | uR[1,4] = 8.18823529 | iN[2,2] = -0.03105882 | uN[2,2] = 12.00000000 |
| iR[1,3] = 0.02964705 | uR[1,5] = 8.18823529 | iN[2,3] = 0.02964705 | uN[2,3] = 0.00000000 |
| iR[1,4] = 0.02047058 | uR[2,1] = 8.89411765 | iN[2,4] = 0.01905882 | uN[2,4] = 8.18823529 |
| iR[1,5] = -0.00141176 | uR[2,2] = 8.18823529 | iN[3,1] = -0.02964705 | uN[3,1] = 8.89411765 |
| iR[2,1] = -0.03105882 | uR[2,3] = 0.00000000 | iN[3,2] = -0.01905882 | uN[3,2] = 12.00000000 |
| iR[2,2] = -0.01905882 | uR[2,4] = 0.00000000 | iN[3,3] = 0.02047058 | uN[3,3] = 0.00000000 |
| iR[2,3] = -0.02964705 | uR[2,5] = 8.89411765 | iN[3,4] = -0.02047058 | uN[3,4] = 8.18823529 |
| iR[2,4] = -0.02047058 | | | |
| iR[2,5] = 0.00141176 | iN[1,1] = -0.00141176 | uN[1,1] = 8.89411765 | u_SRC = 12.00000000 |
| | iN[1,2] = 0.05011764 | uN[1,2] = 12.00000000 | u_GND = 0.00000000 |
| uR[1,1] = 12.00000000 | iN[1,3] = -0.05011764 | uN[1,3] = 0.00000000 | i_SRC = -0.05011764 |
| uR[1,2] = 12.00000000 | iN[1,4] = 0.00141176 | uN[1,4] = 8.18823529 | i_GND = 0.05011764 |

The symbolic solution using fixed values is:
$$i\_GND = u\_SRC * (R_1R_3+R_1R_4+R_1R_5+R_2R_3+R_2R_4+R_2R_5+R_3R_5+R_4R_5) /$$
$$(R_1R_2R_3+R_1R_2R_4+R_1R_2R_5+R_1R_3R_4+R_1R_4R_5+R_2R_3R_4+R_2R_3R_5+R_3R_4R_5)$$

The problems to be solved based on this example are:

**FEOB1**. Numerical solution of the system for fixed values
**FEOB2**. Numerical solution of the system for tolerance 10%, 20% of the resistors
**FEOB3**. Symbolic solution of the system for fixed values
**FEOB4**. Symbolic solution of the system for tolerance 10%, 20% of the resistors
**FEOB5**. Diagnosis as optimization. This requires support for OR-ed constraints (see BLO7)

### 3.2.2 FENB. N Boxes

The constraint system (system of equations) describing the first example with N boxes is:

```
for i=1 to n
{
resistors:
   for j=1 to 5
      i1_Rj_Bi + i2_Rj_Bi = 0; u1_Rj_Bi - u2_Rj_Bi = Rj * i1_Rj_Bi;
nodes:
   for j=1 to 4
      i1_Nj_Bi+i2_Nj_Bi+i3_Nj_Bi = 0; u1_Nj_Bi = u2_Nj_Bi;u1_Nj_Bi = u3_Nj_Bi;
wires:
      i1_N1_Bi + i2_R5_Bi = 0; u1_N1_Bi = u2_R5_Bi;
      i2_N1_Bi + i2_R1_Bi = 0; u2_N1_Bi = u2_R1_Bi;
      i3_N1_Bi + i1_R3_Bi = 0; u3_N1_Bi = u1_R3_Bi;
      i2_N2_Bi + i1_R1_Bi = 0; u2_N2_Bi = u1_R1_Bi;
      i3_N2_Bi + i1_R2_Bi = 0; u3_N2_Bi = u1_R2_Bi;
      i2_N3_Bi + i2_R3_Bi = 0; u2_N3_Bi = u2_R3_Bi;
      i3_N3_Bi + i2_R4_Bi = 0; u3_N3_Bi = u2_R4_Bi;
      i1_N4_Bi + i1_R5_Bi = 0; u1_N4_Bi = u1_R5_Bi;
      i2_N4_Bi + i2_R2_Bi = 0; u2_N4_Bi = u2_R2_Bi;
      i3_N4_Bi + i1_R4_Bi = 0; u3_N4_Bi = u1_R4_Bi;
}
source:
      u_SRC = 12;
ground:
      u_GND = 0;
SRC & GND wires:
      i_SRC + i1_N2_B1 = 0; u_SRC = u1_N2_B1;
      i1_N3_Bn + i_GND = 0; u1_N3_Bn = u_GND;
Boxes connections:
for i=2 to n
      i1_N3_B_{i-1} + i1_N2_Bi = 0; u1_N3_B_{i-1} = u1_N2_Bi;
```

The symbolic solution for i_GND, using fixed values is:

$$i\_GND = (u\_SRC / N) * (R_1R_3+R_1R_4+R_1R_5+R_2R_3+R_2R_4+R_2R_5+R_3R_5+R_4R_5) /$$
$$(R_1R_2R_3+R_1R_2R_4+R_1R_2R_5+R_1R_3R_4+R_1R_4R_5+R_2R_3R_4+R_2R_3R_5+R_3R_4R_5)$$
$$= (u\_SRC / N) * (71 / 17000)$$

The problems to be solved based on this example are:

**FENB1**. Numerical solution of i_GND for fixed values, N=2,3,4,5,10,20,40,80,100,200,500

**FENB2**. Numerical solution of i_GND for tolerance 10%, 20% of the resistors, N=2,3,4,5,10,20,40,80,100,200

**FENB3**. Symbolic solution of i_GND for fixed values, N=2,3,4,5,10,20,40,80,100,200,500

**FENB4**. Symbolic solution of i_GND for tolerance 10%, 20% of the resistors, N=2,3,4,5,10,20,40,80,100,200

**FENB5**. Diagnosis as optimization. This requires support for OR-ed constraints (see BLO7)

### 3.3 Solving the Second Example (SE)

#### 3.3.1 SEOB. One Box

The constraint system describing the second example with just one box is:

```
resistors:
   for j in {2,3,5}
       i1_Rj + i2_Rj = 0; u1_Rj - u2_Rj = Rj * i1_Rj;
diodes:
   for j in {1,4}
       i1_Dj + i2_Dj = 0;((u1_Dj=u2_Dj and i1_Dj>=0)or(u1_Dj<u2_Dj and i1_Dj=0));
nodes:
   for j=1 to 4
       i1_Nj + i2_Nj + i3_Nj = 0; u1_Nj = u2_Nj; u1_Nj = u3_Nj;
wires:
       i1_N1 + i2_R5 = 0; u1_N1 = u2_R5;
       i2_N1 + i2_D1 = 0; u2_N1 = u2_D1;
       i3_N1 + i1_R3 = 0; u3_N1 = u1_R3;
       i2_N2 + i1_D1 = 0; u2_N2 = u1_D1;
       i3_N2 + i1_R2 = 0; u3_N2 = u1_R2;
       i2_N3 + i2_R3 = 0; u2_N3 = u2_R3;
       i3_N3 + i2_D4 = 0; u3_N3 = u2_D4;
       i1_N4 + i1_R5 = 0; u1_N4 = u1_R5;
       i2_N4 + i2_R2 = 0; u2_N4 = u2_R2;
       i3_N4 + i1_D4 = 0; u3_N4 = u1_D4;
source:
       u_SRC = 12;
ground:
       u_GND = 0;
SRC & GND wires:
       i_SRC + i1_N2 = 0; u_SRC = u1_N2;
       i1_N3 + i_GND = 0; u1_N3 = u_GND;
```

The symbolic solution using fixed values is: i_GND = u_SRC / R2

The problems to be solved based on this example are:

**SEOB1**. Numerical solution of the system for fixed values

**SEOB2**. Numerical solution of the system for tolerance 10%, 20% of the resistors

**SEOB3**. Symbolic solution of the system for fixed values

**SEOB4**. Symbolic solution of the system for tolerance 10%, 20% of the resistors

**SEOB5**. Diagnosis as optimization. This requires support for OR-ed constraints (see BLO7)

### 3.3.2 SENB. N Boxes

The constraint system describing the second example with N boxes is:

```
for i=1 to n
{
resistors:
   for j in {2,3,5}
       i1_Rj_Bi + i2_Rj_Bi = 0; u1_Rj_Bi - u2_Rj_Bi = Rj * i1_Rj_Bi;
diodes:
   for j in {1,4}
       i1_Dj_Bi+i2_Dj_Bi=0;
       ((u1_Dj_Bi=u2_Dj_Bi and i1_Dj_Bi>=0)or(u1_Dj_Bi<u2_Dj_Bi and i1_Dj_Bi=0));
nodes:
   for j=1 to 4
       i1_Nj_Bi+i2_Nj_Bi+i3_Nj_Bi = 0; u1_Nj_Bi = u2_Nj_Bi; u1_Nj_Bi = u3_Nj_Bi;
wires:
       i1_N1_Bi + i2_R5_Bi = 0; u1_N1_Bi = u2_R5_Bi;
       i2_N1_Bi + i2_D1_Bi = 0; u2_N1_Bi = u2_D1_Bi;
       i3_N1_Bi + i1_R3_Bi = 0; u3_N1_Bi = u1_R3_Bi;
       i2_N2_Bi + i1_D1_Bi = 0; u2_N2_Bi = u1_D1_Bi;
       i3_N2_Bi + i1_R2_Bi = 0; u3_N2_Bi = u1_R2_Bi;
       i2_N3_Bi + i2_R3_Bi = 0; u2_N3_Bi = u2_R3_Bi;
       i3_N3_Bi + i2_D4_Bi = 0; u3_N3_Bi = u2_D4_Bi;
       i1_N4_Bi + i1_R5_Bi = 0; u1_N4_Bi = u1_R5_Bi;
       i2_N4_Bi + i2_R2_Bi = 0; u2_N4_Bi = u2_R2_Bi;
       i3_N4_Bi + i1_D4_Bi = 0; u3_N4_Bi = u1_D4_Bi;
}
source:
       u_SRC = 12;
ground:
       u_GND = 0;
SRC & GND wires:
       i_SRC + i1_N2_B1 = 0; u_SRC = u1_N2_B1;
       i1_N3_Bn + i_GND = 0; u1_N3_Bn = u_GND
Boxes connections:
for i=2 to n
       i1_N3_B_{i-1} + i1_N2_Bi = 0; u1_N3_B_{i-1} = u1_N2_Bi;
```

The symbolic solution for i_GND, using fixed values is: i_GND = (u_SRC / N) / R2.
The problems to be solved based on this example are:
**SENB1**. Numerical solution of i_GND for fixed values, N=2,3,4,5,10,20,40,80,100,200,500
**SENB2**. Numerical solution of i_GND for tolerance 10%, 20% of the resistors,
   N=2,3,4,5,10,20,40,80,100,200
**SENB3**. Symbolic solution of i_GND for fixed values, N=2,3,4,5,10,20,40,80,100,200,500
**SENB4**. Symbolic solution of i_GND for tolerance 10%, 20% of the resistors,
   N=2,3,4,5,10,20,40,80,100,200
**SENB5**. Diagnosis as optimization. This requires support for OR-ed constraints (see BLO7)

## 4 Experimental Results

It is obvious that for solving the above problems an ideal solver should be able to:
- handle real linear and nonlinear constraints
- handle interval real constraints and interval computations
- support symbolic processing
- support constrained global optimization

Since the constraint sets use indexed variables it would be very useful if the solver also supports variable indexing, otherwise we would have to generate all the individual constraints.

*Unfortunately, there is no current solver able to satisfy all these conditions to our knowledge.*

Therefore we will try to solve the various problems formulated above using various systems that are suitable. We tried to solve the proposed problems using the following systems:

- **OPL Studio** – a world leading modeling tool from ILOG
- **AMPL** – a well known modeling language able to interact with numerous solvers
- **GAMS** – another modeling language
- **Mathematica** – a famous Computer Algebra System from Wolfram Research
- **UniCalc** – a system developed at the Russian Institute for Artificial Intelligence

## *4.1 Baby Example*

The Baby Example (BE) is a very simple example we used to test various features of the benchmarked systems. The table below shows what features were supported by each of the benchmarked systems.

| Problem | Features tested | OPL | AMPL | GAMS | Mathematica | UniCalc |
|---|---|---|---|---|---|---|
| BLNS1 | linear numerical solving, fixed values | x | x |  | x |  |
| BLNS2 | linear numerical solving, interval values |  |  |  | x |  |
| BLNS3 | linear numerical solving, OR-ed fixed values |  |  |  | x |  |
| BLNS4 | linear numerical solving, OR-ed interval values |  |  |  | x |  |
| BNNS1 | nonlinear numerical solving, fixed values |  | x |  | x |  |
| BNNS2 | nonlinear numerical solving, interval values |  |  |  | x |  |
| BNNS3 | nonlinear numerical solving, OR-ed fixed values |  |  |  | x |  |
| BNNS4 | nonlinear numerical solving, OR-ed interval values |  |  |  | x |  |
| BSS1 | symbolic solving, fixed values |  |  |  | x |  |
| BSS2 | symbolic solving, interval values |  |  |  | x |  |
| BSS3 | symbolic solving, OR-ed fixed values |  |  |  | x |  |
| BSS4 | symbolic solving, OR-ed interval values |  |  |  | x |  |
| BLO1 | simple linear optimization | x | x | x | x |  |
| BLO2 | interval computations & linear optimization |  |  |  |  |  |
| BLO3 | interval computation simulation by linear optimization | x | x | x | x |  |
| BLO4 | strict inequalities |  |  |  | x |  |
| BLO5 | direct support for OR-ed constraints, linear optimization |  |  |  |  |  |
| BLO6 | simulation of OR-ed constraints |  |  |  |  |  |
| BLO7 | diagnosis as optimization |  |  |  |  |  |
| BNO1 | simple nonlinear optimization |  | x |  |  |  |
| BNO2 | interval computations & nonlinear optimization |  |  |  |  |  |
| BNO3 | interval computation simulation by nonlinear optimization |  | x |  |  |  |
| BNO4 | direct support for OR-ed constraints, nonlinear optimization |  |  |  |  |  |

*Table 1. Features supported when solving the Baby Example*

## 4.2 First Example

The First Example (FE) is a more complex example that requires constraint solving and indexing of variables. The following table shows the results for the benchmarked systems.

| Problem | Features tested | OPL | AMPL | GAMS | Mathematica | UniCalc |
|---------|-----------------|-----|------|------|-------------|---------|
| FEOB1 | linear numerical solving, fixed values, one box | x | x | x | x | |
| FEOB2 | linear numerical solving, interval values, one box | | | | x | |
| FEOB3 | symbolic solving, fixed values, one box | | | | x | |
| FEOB4 | symbolic solving, interval values, one box | | | | x | |
| FEOB5 | diagnosis as optimization, one box | | | | | |
| FENB1 | linear numerical solving, fixed values, n boxes | x | x | x | x | |
| FENB2 | linear numerical solving, interval values, n boxes | | | | | |
| FENB3 | symbolic solving, fixed values, n boxes | | | | | |
| FENB4 | symbolic solving, interval values, n boxes | | | | | |
| FENB5 | diagnosis as optimization, n boxes | | | | | |

*Table 2. Features supported when solving the First Example*

The following table shows, for the problem FENB1, the solutions for the ground current (i_GND) for various values of "n". Here "NA" means "not available" due to the limitations of the demo versions of AMPL and GAMS, and "∞" means that the computation took a very long time, too long for any practical purposes.

| n | Exact value | OPL | AMPL | GAMS | Mathematica |
|---|-------------|-----|------|------|-------------|
| 1 | 0.05011765 | 0.05011764 | 0.05011765 | 0.05011765 | 0.0501176 |
| 2 | 0.02505882 | 0.02505882 | 0.02505882 | 0.02505882 | 0.0250588 |
| 3 | 0.01670588 | 0.01670588 | 0.01670588 | 0.01670588 | 0.0167059 |
| 4 | 0.01252941 | 0.01252941 | 0.01252941 | 0.01252941 | 0.0125294 |
| 5 | 0.01002353 | 0.01002352 | 0.01002353 | 0.01002353 | 0.0100235 |
| 10 | 0.00501176 | 0.00501176 | NA | NA | inf |
| 20 | 0.00250588 | 0.00250588 | NA | NA | inf |
| 40 | 0.00125294 | 0.00125294 | NA | NA | inf |
| 80 | 0.00062647 | 0.00062647 | NA | NA | inf |
| 100 | 0.00050118 | 0.00050117 | NA | NA | inf |
| 200 | 0.00025059 | 0.00025058 | NA | NA | inf |
| 500 | 0.00010024 | 0.00010023 | NA | NA | inf |

*Table 3. Values of i_GND for the FENB1 Problem*

The table below summarizes the execution time for FENB1 for various values of "n". We used an Intel Celeron (550 MHz) based system running Windows 98.

| n | Constraints | Variables | Time (sec) OPL | Time (sec) AMPL (MINOS solver) | Time (sec) GAMS (CPLEX solver) | Time (sec) MATHEMATICA |
|---|---|---|---|---|---|---|
| 1 | 48 | 48 | 0.01 | 0.05 | 0.00 | 2.14 |
| 2 | 92 | 92 | 0.01 | 0.05 | 0.00 | 13.63 |
| 3 | 136 | 136 | 0.01 | 0.06 | 0.05 | 44.88 |
| 4 | 180 | 180 | 0.01 | 0.11 | 0.05 | 110.07 |
| 5 | 224 | 224 | 0.01 | 0.11 | 0.00 | 230.19 |
| 10 | 444 | 444 | 0.02 | NA | NA | inf |
| 20 | 884 | 884 | 0.06 | NA | NA | inf |
| 40 | 1764 | 1764 | 0.17 | NA | NA | inf |
| 80 | 3524 | 3524 | 0.28 | NA | NA | inf |
| 100 | 4404 | 4404 | 0.44 | NA | NA | inf |
| 200 | 8804 | 8804 | 0.88 | NA | NA | inf |
| 500 | 22004 | 22004 | 2.53 | NA | NA | inf |

*Table 4. Execution time for the FENB1 Problem*

In the above table, we marked with NA the "not available" values due to the limitations of the demo versions of AMPL and GAMS we used, and " inf " means that the computation took a very long time, too long for any practical purposes.

Notice the abnormal behavior of the GAMS system, which shows strange results like a null execution time for n = 5. It should also be considered that while OPL and AMPL gave the same results for any runs for the same "n", GAMS gave totally unpredictable results and Mathematica gave results around the values in the Table 4.

### *4.3 Second Example*

None of the tested systems was able to solve any of the problems SEOB1-5, SENB1-5.

## 5 Conclusions

### *5.1 Features supported*

In this paragraph we analyze the benchmarked systems from the point of view of the features they support (see Table 1 and 2). We shall start with **UniCalc**, who failed to solve any of the proposed problems, so even if theoretically it supports say interval computations, we have to conclude that it is a totally inappropriate system for solving problems from the BE, FE or SE classes.

While UniCalc showed the poorest behavior, from the Table 1 we can see that **Mathematica** seems to have the best behavior solving most of the proposed problems. It is noticeable its ability to handle symbolic computations and interval computations, unsurprisingly expected from a Computer Algebra System. However we must make a few remarks here.

First, Mathematica is in fact unable to handle constraint solving problems and it is just a pure coincidence that most of the constraint systems to be solved were in fact *systems of equations*, which Mathematica handles very well; were there a single inequality in those systems and Mathematica would have failed.

Second, Mathematica is very weak in the optimization field; it can only handle linear systems and has a very strong constraint on variables: they must all be positive. Since in practice you can't know from the beginning whether the variables are positive or not, this makes Mathematica totally inappropriate for any practical optimization problems.

**OPL, AMPL** and **GAMS** are somehow similar in the sense that they are all modeling languages; their power is in the solvers they come with, although the power of the language itself is an intrinsic feature. The solvers used are CPLEX, MINOS and CPLEX respectively.

We consider **GAMS** the weakest of the three languages; it can't handle constraint solving, it can handle just optimization. It has a very difficult syntax, it was the hardest to learn because of its "free style syntax" which gives too much liberty and makes very difficult to distinguish between the program and the comments; it has very intriguing syntactic features, like using =e=, =l=, =g= for the well known "equal", "less than or equal" and "greater than or equal" operators; programs are not parametrizable except for command line parameters; the output is extremely verbose; the error messages are ambiguous.

**OPL** is very easy to learn and use due to its excellent windowed IDE. It can handle constraint solving and optimization but unfortunately nonlinear constraints over reals are not supported nor is there any attempt to reduce them to linear constraints. It proved to be the fastest system we tested.

**AMPL** is very similar to OPL (in fact the OPL syntax was inspired from AMPL) but it doesn't come with any IDE which makes it a little bit difficult to use. It has the advantage that it can be used with many external solvers. Before solving a system, it applies a "presolve" stage that uses some transformations on the original system trying to make it simpler; this is how simple nonlinear problems like BNNS1 are reduced to linear ones and solved.

## 5.2 Accuracy

As shown in the Table3, all of the systems we tested (except UniCalc) were very accurate, giving values identical to the theoretically computed values or differing at the last decimal digit. It is remarkable that OPL, AMPL and GAMS support user defined accuracy while Mathematica doesn't.

## 5.3 Speed

As shown in the Table 4, the slowest system is Mathematica and the fastest is OPL. AMPL (using MINOS) is faster than Mathematica but slower than OPL. We cannot say anything about GAMS since it shows confusing values and we didn't have a commercial version to be able to run large problems.

## 5.4 Final Remarks

Finally we may conclude saying that for the BE, FE and SE classes of problems, **Mathematica** is the best system to use, AMPL and OPL are next followed by GAMS and UniCalc is last being unable to solve any of the proposed problems.

Although **Mathematica** seems to be the best for the proposed problems, it is in fact **unable to handle constraint solving problems** and it is just a pure coincidence that most of the constraint systems to be solved were in fact systems of equations, which Mathematica handles very well; were there a single inequality in those systems and Mathematica would have failed. It is also important to note that Mathematica was the slowest of the tested systems.

Therefore, we consider **ILOG OPL** to be the system of choice if **speed** is a critical factor, although **AMPL** solves a **larger range of problems**.

It is interesting to notice that no system was able to solve the SE class of problems and other problems that involved "OR"-ed constraints (such as diagnosis as optimization problems). However, private communications with the members of the AMPL and GAMS teams show that such a feature is desired and will be introduced in future versions of these systems.